\documentclass[11pt]{article}
\pdfoutput=1
\usepackage{graphicx}
\usepackage{mathtools}
\usepackage{amssymb}

\usepackage[T1]{fontenc}
\usepackage{fancyhdr}
\fancypagestyle{firstpage}{\fancyhf{}\fancyhead[R]{\includegraphics[height=1in,
    keepaspectratio=true]{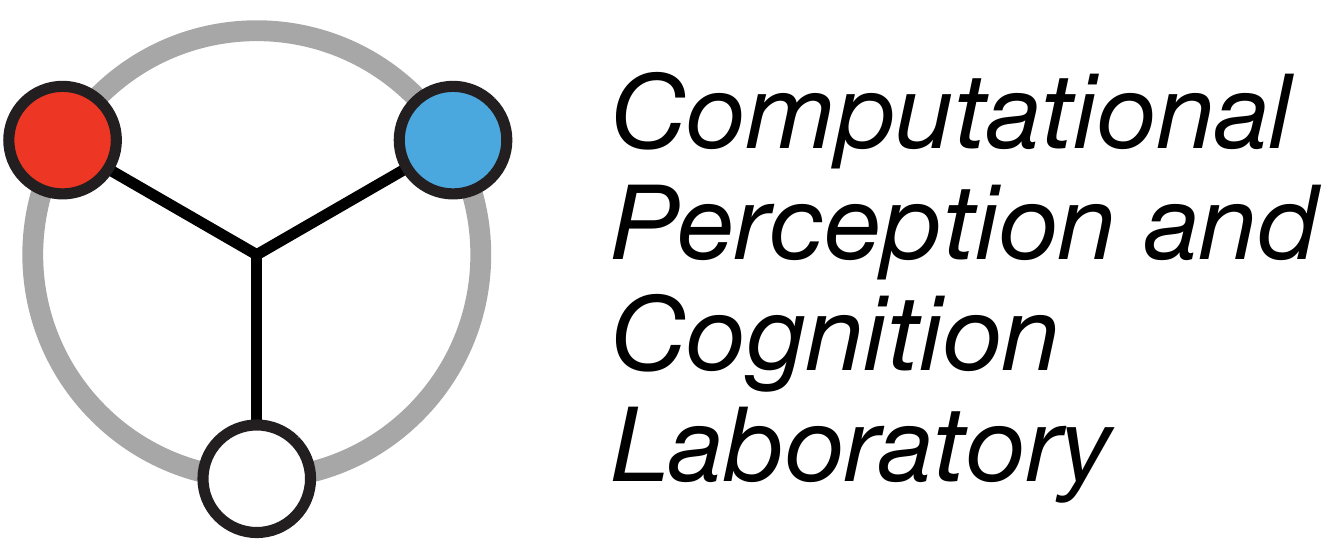}}\fancyfoot[L]{Philadelphia, \today}}
\fancypagestyle{plain}{\fancyhf{}}

\usepackage{geometry}
\newcommand{\m}{{\bf m}}

\long\def\comment#1{}

\begin{document}

% COVERPAGE
\thispagestyle{firstpage}
\begin{tabbing}
  MANUSCRIPT\hspace{40mm}\=\\[40mm]
  Authors:\>Alexander Tank and Alan A. Stocker\\
  \> \\
  Affiliation: \>Department of Psychology and\\
  \>Department of Electrical and Systems Engineering\\
  \>University of Pennsylvania\\
  \>\\
  Correspondence: \>Dr. Alan A. Stocker\\
  \>Computational Perception and Cognition Laboratory\\
  \>3401 Walnut Street 313C\\
  \>Philadelphia, PA 19104-6228\\
  \>U.S.A.\\[2ex]
  \>astocker@sas.upenn.edu\\
  \>phone: +1 215 573 9341\\
  \>\\
  Journal: \>arXiv\\
  \>\\
  Classification: \>Quantiative Biology (Neurons and Cognition)\\
  \>\\
 \comment{ Statistics: \>xx pages, xx figures\\}
\end{tabbing}
\newpage

\ \\
% extra empty page
\thispagestyle{empty}
\newpage

% MANUSCRIPT
\pagestyle{plain}

% title
\section*{\LARGE Biased perception leads to biased action: Validating a Bayesian model of interception}

{\large Alexander Tank and Alan A. Stocker}

% layout
% title
% abstract
% intro
% results
% - new method
% -- setup
% -- trial allocation
% -- observer model
% - experimental validation
% -- 2AFC can be biased
% discussion

% for both detection and discrimination threshold measurements.

\section*{Abstract} 
 We tested whether and how biases in visual perception might influence motor actions. To do so, we designed an
  interception task in which subjects had to indicate the time when a moving object, whose trajectory was occluded,
  would reach a target-area. Subjects made their judgments based on a brief display of the object's initial motion at a
  given starting point. Based on the known illusion that slow contrast stimuli appear to move slower than high contrast
  ones, we predict that if perception directly influences motion actions subjects would show delayed interception times
  for low contrast objects. In order to provide a more quantitative prediction, we developed a Bayesian model for the
  complete sensory-motor interception task. Using fit parameters for the prior and likelihood on visual speed from a
  previous study we were able to predict not only the expected interception times but also the precise characteristics
  of response variability. Psychophysical experiments confirm the model's predictions. Individual differences in
  subjects' timing responses can be accounted for by individual differences in the perceptual priors on visual
  speed. Taken together, our behavioral and model results show that biases in perception percolate downstream and cause
  action biases that are fully predictable.

\section{Motivation}
Bayesian models of perceptual inference explain many perceptual biases~\cite{Knill07, alan2006, hedges2011}. By
leveraging prior knowledge, Bayesian inference provides a principled and optimal strategy for an observer to infer the
state of a perceptual variable from observed noisy evidence. Perceptual biases arise in this context because prior
beliefs about the environmental statistics influence the perceptual process.  Well documented examples include biases
cue combination \cite{Knill07}, low contrast biases in perceived speed \cite{alan2006}, and biases towards the cardinals
in orientation perception \cite{Odelia06, Girshick11}, to name a few. A separate line of work suggests that many visual
illusions fail to lead to equivalent biases in motor behavior \cite{GooddaleREV}.  For example, the Ponzo illusion,
where lines of the same length are perceived differently due to receding distance, does not translate into differences
in motor commands for grasping these lines \cite{Ganel08}. From a Bayesian perspective, the disconnect between
perception and action appears puzzling. If a perceptual illusion can be explained in the context of optimal Bayesian
inference, then we would expect the illusion to translate into biased behavior because the Bayesian framework is
grounded in the statistics of the physical world \cite{Giesler03}.  Furthermore, why allocate costly resources for
optimal perception if these perceptions are not used in guiding actions?
 
Speed perception has been extensively studied within the Bayesian framework \cite{alan2006} and provides an entry point
for dissecting the influence of Bayesian perception upon action. Speed perception is traditionally studied with a two
alternative forced choice task (2AFC), where subjects must choose the faster of two motion stimuli. Under this setup,
low contrast stimuli are perceived to move slower relative to high contrast stimuli of the same speed \cite{Thompson81}.
A prior distribution favoring slower speeds combined with a wider likelihood width for low contrast qualitatively
explains the slow speed illusion and provides a tight fit to 2AFC data \cite{alan2006, hedges2011}. 

\begin{figure}
\begin{center}
  \includegraphics[width=\linewidth]{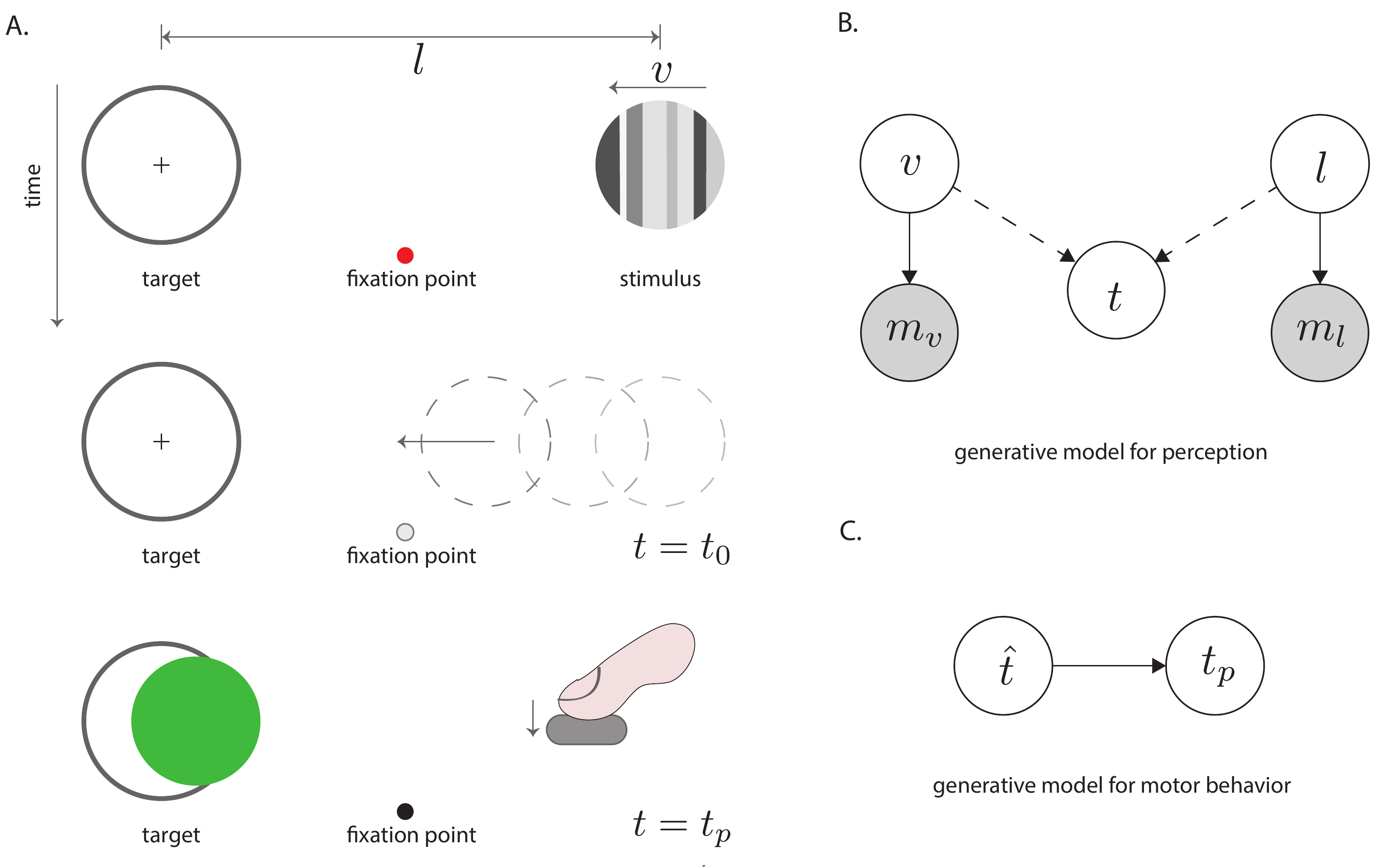}
  \caption{{\bf Interception task: task structure and generative model.}  (A) (\emph{top}) Subjects fixate on the small
    red dot on the bottom of the screen while viewing a drifting motion stimulus \cite{alan2006} for 1s at an
    eccentricity of $6^{\circ}$. (\emph{middle}) At the moment the motion stimulus disappears, the fixation dot turns
    white to alert the subjects that the stimulus has been released. The stimulus moves, occluded, at a constant speed
    across the screen.  (\emph{bottom}) Subject makes a button press to indicate when the stimulus would have reach the
    center of the target area. For medium contrast conditions, feedback was provided indicating the end position of the
    stimulus at the time of the bottom press. Position was shown in green if the stimulus is within the target area and
    in red otherwise. (B) The generative model for perception shown as a directed graph. The values of the latent
    physical variables, stimulus velocity, $v$, and length, $l$, are combined using a simple physics model
    (deterministic) to define a time, $t$, for the stimulus to reach the target area. Observed variables represent the
    sensory measurements for speed, $m_v$, and length, $m_l$. The width of the noise distribution for $m_v$ is contrast
    dependent and changes based on the contrast from trial to trial. (C) The generative model for the motor control
    assumes that the time of the key press (action), $t_p$, is drawn from a Gaussian distribution with mean equal to the
    perceptual time estimate, $\hat{t}$. Solid arrows indicate stochastic dependencies whereas dotted lines indicate
    deterministic dependencies.}
\label{fig:task}
\end{center}
\end{figure}

We developed a sensory-motor interception task in order to assess whether or not these perceptual speed biases translate to
biases in timing actions. In this interception task subjects briefly view a moving grating stimuli contained in a
circular window (Figure \ref{fig:task}).  The stimuli then disappears at which point it moves occluded with constant
velocity across the screen. Subjects are instructed to make a button press when they believe the stimuli has crossed to
the other side of the screen and coincides with some target a fixed distance away from stimulus presentation.
Importantly, the moving stimuli are of differing contrasts to induce the slow speed illusion.  We also present different
speeds to obtain timing data for multiple speed contrast pairs. To formalize a clear hypothesis, we developed a Bayesian
model for this task that assumes that a subject's timing response is optimal in order to best most accurately intercept
the object in the target area. Such an optimal model necessarily predicts that the perceptual biases (because they are
optimal) translate to the corresponding motor biases.

In what follows we first derive the Bayesian model for our the described interception task.  We then explain how we
perform the task with human subjects and present the measured behavioral results. Finally we compare the data to
predictions of our model based on prior and likelihood parameters obtained from psychophysical experiments in a previous
study of speed perception.

\section{Interception model}
We assume each subject observes both the speed of the moving stimulus and the distance between the stimulus and the target (see
Fig.~\ref{fig:task}A).  These separate perceptual measurements must be integrated and decoded in some sensible way in order to determine an
adequate time estimate.  We formalize perception, intuitive physics, and action in our task using generative models and then derive an
optimal time estimator based on assumptions of rationality, the independence assumptions of our perceptual model, and the motor noise in
timing actions.

\subsection{Generative model for perception}
We model each trial independently, and in what follows we avoid trial subscripts for convenience. On each trial the
stimulus speed, $v$, and length between stimulus center and target, $l$, are each drawn independently from their
respective natural prior distributions, $p(v)$ and $p(l)$. The time, $t$, the stimulus needs to reach the center
of the target is dependent only on the physical speed and length and is thus drawn from the conditional distribution,
$p(t|v,l)$.  Speed, length, and time are the \emph{latent} variables that we combine in a vector ${\bf h} = \{v, l, t\}$.
The observable variables ${\bf m} = \{m_v, m_l\}$, are the noisy sensory measurements of the true speed
and length and are each drawn independently from their conditional distributions, $p(m_v|v)$ and $p(m_l|l)$.
Taken together, these conditional and prior distributions define a generative model over the observable variables, ${\bf
  m}$, and latent physical variables, ${\bf h}$. The joint distribution for each trial becomes:
$$p({\bf m}, {\bf h}) = p(v)p(l)p(m_v|v)p(m_l|l)p(t|v,l)$$
and is displayed graphically in Figure \ref{fig:task}. The observer has access to only the measurements and must infer
the physical state.  The posterior distribution of the physical variables given the measurements is:
$$p({\bf h}|{\bf m}) = \frac{p(m_v|v)p(v)p(m_l|l)p(l)p(t|v,l)}{p(m_v)p(m_l)} \nonumber \\
= p(v|m_v) p(l|m_l)p(t|v,l)$$ where the marginal $p(m_v,m_l)$ = $p(m_v)p(m_l)$ because the physical sources of the
measurements are independent.  Furthermore, we see that the marginal distribution for latent physical variables is
separable into independent posterior distributions for speed and length and the conditional distribution for time.

\subsection{Physics model}
We embed the physical relationship between time, velocity and length in the conditional distribution $p(t|v,l)$. If we
assume that length and velocity are constant, rather than fluctuating due to some unknown latent forces, the conditional
density becomes deterministic:
$$p(t|v,l) = \delta(t,q(v,l))$$
where $\delta$ is the Dirac delta measure that assigns all of its probability mass to the second argument. The function
$q(v,l)$ is the \emph{physics model} which maps speed and length to time.  Because we assume constant speed and length,
the physics model is the simple Newtonian relationship: $q(v,l) = \frac{l}{v}$.

\subsection{Action model}
The timing decision, $\hat{t}({\bf m})$, is a deterministic function of both length and speed measurements.  We assume
the subject has only \emph{indirect control} over its motor commands, in the sense that the produced motor output,
$t_p$, stochastically depends on the perceptual estimate $\hat{t}(\m)$ via the distribution $p(t_p|\hat{t}(\m))$ (Figure
\ref{fig:task}). We model $p(t_p|\hat{t}(\m))$ as a Gaussian distribution with mean $\mu_p = \hat{t}(\m)$ and
standard deviation that grows proportionally with the mean, $\sigma_p = w_p \mu_p$ where $w_p$ is the time production
Weber fraction \cite{JazayeriNat}.

\subsection{Estimator of travel time}
We model timing action choice in our task using decision theory \cite{JuliaCOG}, which allows us to identify an optimal
estimator from a class of functions, $f({\bf m})$, that maps the set of measurements directly onto a timing decision. We
assume subjects operate rationally and try to accrue as much task dependent reward as possible. Thus our optimal
estimator, $\hat{t}({\bf m})$, is one which minimizes the \emph{posterior loss} between the timing actions, $t_p$, that
depend stochastically on the timing estimate, and the latent physical variables:
$$\hat{t} ({\bf m}) = \underset{f({\bf m})}{\operatorname{argmin}}\int \mathcal{L}({\bf h},f({\bf m})) p({\bf h} | {\bf m}) \, d {\bf h}$$
where $\mathcal{L}$ is the the expected penalty for making the timing decision $f({\bf m})$ in physical state ${\bf h}$
and $p({\bf h}|\bf{m})$ is the posterior distribution of our generative model.  Of course, in our model the timing
actions depend stochastically on the timing decision so $\mathcal{L}$ is determined by integrating over all possible
timing actions:
$$\mathcal{L}({\bf h}, f(m)) = \int L({\bf h},t_p) p(t_p|f(m)) \, d t_p$$ 
where $L({\bf h},t_p)$ is the exact loss for performing action $t_p$ in latent state ${\bf h}$ and $p(t_p|f({\bf m}))$ is the probability of performing motor command $t_p$ under time decision $f({\bf m})$, as defined above.

%\subsubsection{Loss Function}
While interception actions are performed in time, actions are chosen in reference to the distance moved by the
object. We thus decompose the loss function, $L$, into two separate functions that measure loss along different
dimensions:
$$L({\bf h}, t_p)= L_L({\bf h}, t_p) + \eta  L_T({\bf h}, t_p)$$
Where $L_L$ measures loss in distance, $L_T$ measures loss in time, and $\eta$ parameterizes the weight of loss in time
with respect to loss in distance.  A natural form for loss in distance is the squared error between the true length,
$l$, and the distance traveled at time press $t_p$:
$$L_L({\bf h}, t_p) = \big(l - v t_p \big)^2 $$
where $v t_p$ is the distance traveled by the stimulus moving at speed $v$ at time $t_p$. This loss function captures
the motivational thrust of our task: to land the moving stimulus as close to the target center as possible.  We also
assume that subjects prefer shorter timing actions to finish the experiment quickly.  The loss function in time then
takes a linear form:
$$L_T({\bf h}, t_p) = t_p$$
Under these loss functions $\eta$ controls the trade off between task performance and completion time.
%\subsubsection{Optimal Time Estimator}
With these model specifications, one can show that the optimal estimator that minimizes the posterior loss is given by:
$$\hat{t}({\bf m}) = f_{opt} ({\bf m}) = \frac {E (l|m_l)  E (v|m_v) - \frac{\eta}{2}}{(w_p^2 + 1)  E (v^2|m_v)} $$
where $E(l |m_l)$ and $E(v|m_v)$ are the expectations of the posterior distributions for $l$ and $v$ respectively and
$w_p$ is the timing noise fraction. To derive this expression we plug the full loss function and posterior distribution
into the expected posterior loss and then find the minima of the resulting expression.  We first integrate through each
term:
\begin{align*}
  Loss &= \iint \Big( l^2 - 2 l v t_p + v^2 t_p^2 + \eta t_p\Big) p(t_p|f({\bf m})) d t_p p({\bf h}|{\bf m}) \, d {\bf
    h} \\ &= \int l^2 p(l|m_v) dl - 2 \iiint l v t_p p(t_p|f({\bf m})) p(v|m_v) p(l|m_l) \, dv \, dl \, d t_p \\ & +
  \iint v^2 t_p^2 p(v|m_v) p(t_p|f({\bf m})) \, dv \, dt_p + \int \eta t_p p(t_p|f({\bf m})) d t_p \\ &= E(l^2|m_l) +
  \eta f({\bf m}) - 2 f({\bf m}) E(l|m_l) E(v|m_v) + E(v^2|m_v)(w_p^2 f({\bf m})^2 + f({\bf m})^2)
 \end{align*}  
 where we have used the fact that probability distributions integrate to one and that the first and second moments of a
 distribution are the mean and variance plus mean squared, respectively. We note that the final expression is quadratic
 in $f({\bf m})$ and thus take the derivative of the above equation with respect to $f({\bf m})$, set it equal to zero,
 and solve for $f({\bf m})$ to obtain the desired result. We now describe our experiment and behavioral results before
 returning to model specifications needed to predict behavior.

\section{Psychophysical experiment}
On each trial subjects are presented with a circular patch which contains a horizontal drifting grating. See Figure
~\ref{fig:task} for example stimuli and \cite{alan2006} for stimulus specifics.  The patch is $3^{\circ}$ in radius and
is located $5.2^{\circ}$ to the right of the center of the screen.  Subjects are told to fixate on a small red circle
$3^{\circ}$ below the center of the screen. A large target ring, $4^{\circ}$ in radius, is located $5.2^{\circ}$ to the
left of screen center with a small cross in its center.  These distances were chosen to ensure that the eccentricity
between fixation and stimulus are identical to previous work, $6^{\circ}$. After one second of motion, the grating
disappears and the red fixation circle turns white to indicate that the stimuli trajectory has begun.  Subjects make a
button press when they believe the center of the stimuli coincides with the center of the target (Figure
~\ref{fig:task}).

Subjects first complete two training blocks with feedback at .3 contrast level to adapt to the task. The remaining
blocks contain seven different speeds, equally tiled between 3 to 10 $\frac{deg}{s}$, at three contrast levels, high
(.8), medium (.3), and low (.1). Feedback remains for medium contrast trials.  Three male subjects ages, 23 to 40,
completed the task. Subject 1 completed 48 trails for each contrast speed pair while Subjects 2 and 3 completed 90
trials.
 
To assess the effects of contrast and speed on both the bias and noise in timing actions we first pool trials of the same
contrast speed pair to obtain a sample distribution over timing actions, ${\bf t}_{v,c} =
\{{t_{1:v,c},...,t_{n:v,c}\}}$, where $v$ and $c$ indicate the speed and contrast and $n$ is the number of trials for
each condition. We then compute three sample statistics for each contrast speed pair: the mean, the variance, and the
skewness. By comparing mean estimates we can determine the bias in timing actions due to contrast. The variance provides
a measure of how the variability in motor behavior is affected by the perceptual noise due to contrast.  Skewness,
measured as the third standardized moment, is another statistic we apply to compare model and data characteristics.
Finally, we calculate sample statistics for an average subject by pooling timing actions across subjects. We do this for
both statistical power and for comparison to model predictions with average parameter values.

\begin{figure}
\begin{center}
   \includegraphics[width=\linewidth]{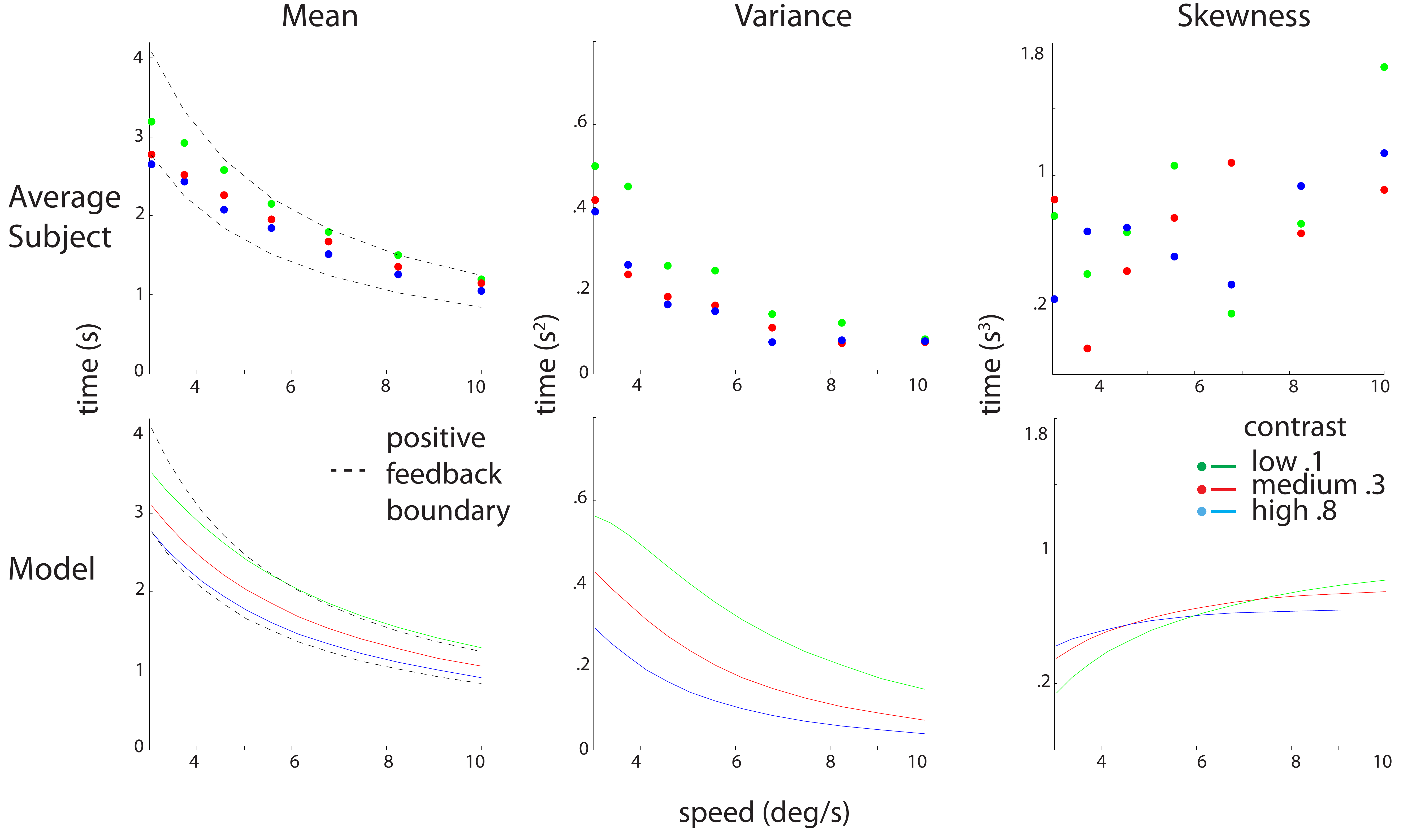} 
   \caption{ {\bf Interception model qualitatively matches average subject data across timing distribution statistics.}
     \emph{(top)} Combined timing data for all three subjects.  For each contrast speed pair the sample mean, sample
     variance, and sample skewness were determined from the pooled data.  \emph{(bottom)} Model timing results using an
     average extracted speed prior distribution and likelihood widths (see \cite{alan2006}). Model and data both display
     a consistent contrast dependent timing bias, whereby lower contrast leads to longer times. Variability is also
     contrast dependent and the model predicts both the general shape and the range of variance values. The model
     reproduces the upward trend in skewness and the range of skewness values. Parameters: $s = .65$ $\eta = 6$.}
 \label{fig:avg}
\end{center}
\end{figure}

%\subsection{Experiment Results}
For the average subject there is significant difference in timing actions ($p \ll .05$) between medium and low contrast
and between medium and high contrast stimuli at all speed levels. Lower contrast stimuli lead to longer times than
medium contrast and medium contrast leads to longer times than high contrast (Figure ~\ref{fig:avg}). Individual subject
data portrays a similar result (Figure ~\ref{fig:datsub}).  Subject 1 shows a significant difference between times for
medium and low contrast at the slower speeds but not for the two fastest tests speeds.  Inspection of the timing plots
shows that there is a contraction in the bias between medium and low contrast at higher speeds. Subject 2 shows no
contraction in bias and maintains significant difference between conditions at higher speeds. For all subjects the
variance decreases as speed increases and is higher for lower contrast than medium and high contrast.  These effects are
reflected in the average subject plots (Figure ~\ref{fig:avg}). For skewness, no clear trend between contrasts is
discernible yet overall skewness increases as speed increases.

\section{Model simulations}
Our interception model is general and can be applied to a variety of tasks. In what follows we adapt this model to predict
behavior in our psychophysical experiment. The model accounts for both the differences in timing bias at low, medium,
and high contrasts and the individual subject differences in bias contraction at high speeds.

As mentioned earlier, the prior and conditional distribution for speed have been previously fit to five different
subjects \cite{alan2006}. We predict behavior for these subjects by plugging in the fit parameter values to our model.
Overall we make behavioral predictions for two of the subjects from the Stocker and Simoncelli 2006 paper, referred to as Subject
1S and Subject 2S, and predictions for an average subject with average parameters taken across all five subjects. We
then compare predicted behavior to human performance by computing the expected value, variance, and skewness of the
simulated timing distributions. We first flesh out the specifics of the prior and conditional distribution for speed.

\subsection{Speed prior and conditional distribution}
While the physical speed distribution across the human retina has yet to be measured, human prior distributions for
speed have previously been extracted from subjects and show a power law shape \cite{alan2006}.  We parameterize this
speed prior by assuming the density in physical space is approximately constant for slow speeds, drops as a power law
for medium to high speeds, and then transitions back to a constant regime at very high speeds (adapted from
\cite{hedges2011}): \vspace{-.1in}$$p(v) \propto \frac {1} {(|v|^2 + b^2)^{d}} + r \vspace{-.0099in}$$%\vspace{-.1in}
where $b$ controls at what speed the density transitions from a constant function of speed to a power law, $d$ controls
the rate of decay, and $r$ controls at what speed the prior transitions back to constant density. We assume this prior
distribution is fixed in advance and is not modulated by the speeds viewed in our experiment. To obtain the parameters
for our simulations we fit this parameterized form of the speed prior to the nonparametric prior distributions extracted
for each of the five subjects.

We now specify an appropriate speed representation for the observer. We choose a mapping function, $f(v)$, which maps
the linear physical speed to a normalized logarithmic speed, $\tilde{v}$ which is in the same space as the perceptual
measurement for speed, $m_v$. The speed transformation is: $ \tilde{v} = f(v) = \log(\frac{v}{v_o} + 1)$ where $v_o$ is
a small normalization constant.  Working in this space allows us to model the conditional speed observation function,
$p(m_v|v)$, as a Gaussian distribution with mean $\tilde{v}$. The standard deviation at each speed is separable into
functions for the contrast of the stimuli, $c$, and speed: $\sigma(c,\tilde{v}) = g(c)h(\tilde{v})$.  The speed and
contrast likelihood parameters are either set to those previously obtained for each subject or linearly interpolated
from these values. Our experiment uses the same distance for every trial so we assume that the subject posterior
distribution over length can be approximated by a delta function, $\delta(l,l_o)$, where $l_o$ is the true distance
between stimuli and target.  The expectation over length becomes $l_o$.

\subsection{Motor adaptation and model output}
Subjects were trained with feedback on stimuli of medium contrast.  Due to the slow speed illusion, this leads them (and our
model) to overestimate the time it takes for the stimuli to reach the target.  After sufficient feedback, timing
behavior for medium contrast shifts to faster, more correct, times.  To model this motor adaptation on a fast time scale
we introduce a scaling factor, $s$, that shrinks the time estimate appropriately to obtain a new corrected time
estimate, $\hat{t}_c (\m)$. Namely, $\hat{t}_c(\m) = s \, \hat{t}(\m)$. Because this adaptation happens quickly it
should only affect motor output and not the estimator, which we assume has been optimized by the observer over a
lifetime of interacting with moving objects and making time judgments.  For simulations the $s$ parameter is chosen
such that the mean timing predictions at medium contrast overlap with the veridical time to contact. The timing noise
fraction, $w_p$, was previously fit to subject timing behavior \cite{JazayeriNat} and we set it to the average of these
maximum likelihood fits, $w_p$ = $.07$.

In order to compare the timing distributions of the data to our model we computed the probability of making action $t_p$ under
speed $v$ by marginalizing out the measurements:
$$p(t_p|v) = \int p(t_p|\hat{t}_c(m_v)) p(m_v|v) d m_v$$
where $p(t_p|\hat{t}_c(m_v))$ and $p(m_v|v)$ are as defined above.  The expected value, variance, and skewness of the timing distribution are all calculated using numerical integration.

\begin{figure}
\begin{center}
\includegraphics[width=\linewidth]{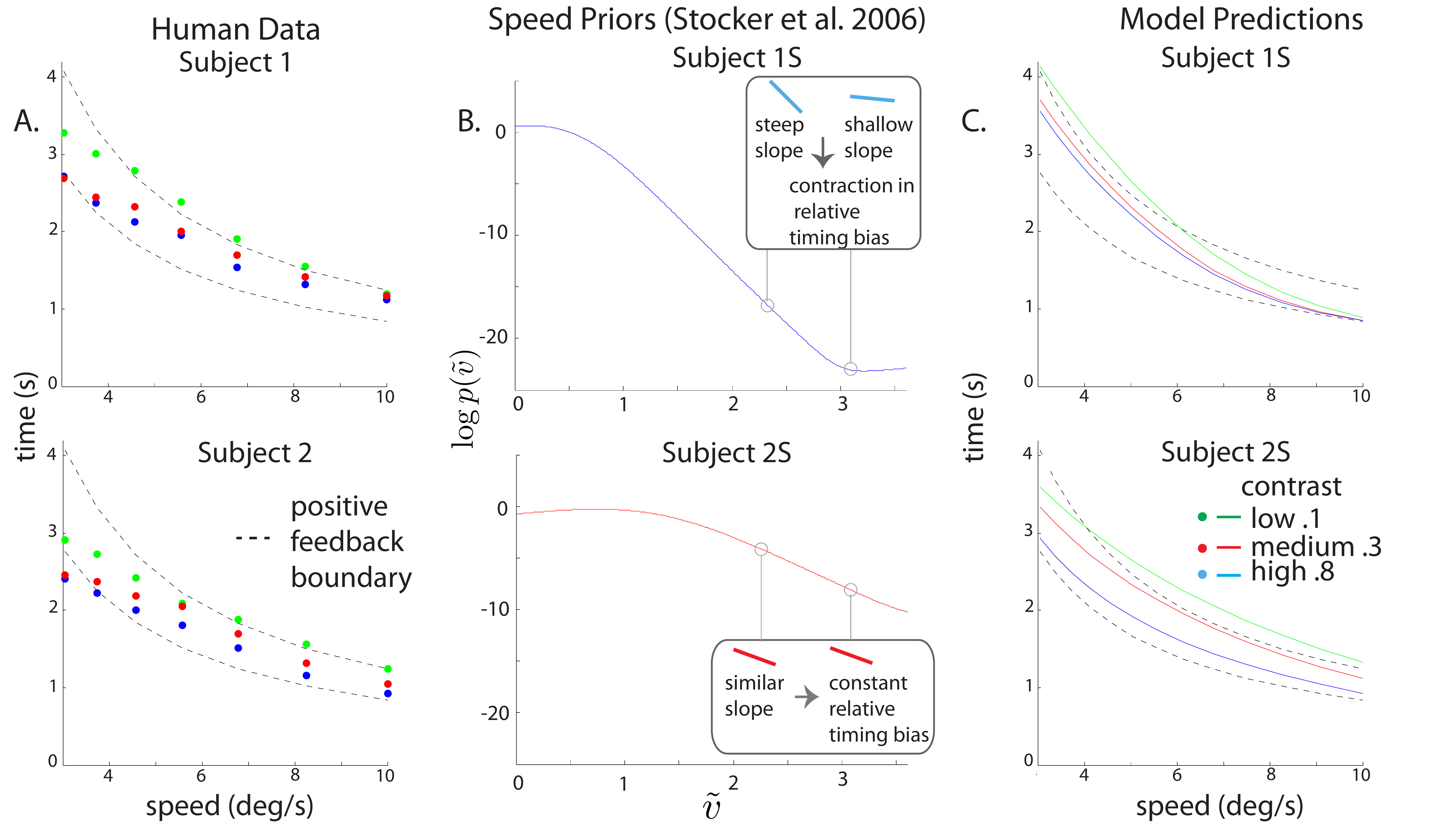}
\end{center}
\caption{{ \bf Variance in previously extracted subject priors explains between subject variance in timing actions.}
  ({A.}) Mean timing actions for Subjects 1 and 2 as functions of speed for each contrast. Subject 2 maintains a roughly
  constant difference in bias between the contrast levels across speeds while Subject 1 shows a contraction in timing
  bias. ({B.}) The extracted parameterized speed prior distribution for both model subjects. The Subject 1S prior slope
  becomes shallow at high speeds while the Subject 2S prior maintains a constant slope. ({C.}) Mean timing model
  predictions for the same two subjects. Again, we see a contraction in timing bias for Subject 1S and not for Subject
  2S which can be partially explained by differences in prior shape. Parameters: 1S $s = .9$ $\eta = 8$ 2S: $s = .8$
  $\eta = 6.4$ }
\label{fig:datsub}
\end{figure}

\subsection{Model results and comparison to data}
The predictions for the average subject show a clear contrast dependent timing bias qualitatively similar to results
from the average human subject.  Furthermore, both model and data show no contraction at higher speeds. The model
variance predictions across contrasts show a steep initial decline followed by a leveling off at higher speeds.  Lower
contrasts induce more variability in timing response, as expected due to the wider width of the conditional distribution
for speed measurements.  This prediction is also qualitatively similar to our human data and within the same variance
range.  At slow speeds lower contrast stimuli predict smaller skewness while as the speeds increase we see a gradual
cross over whereby at high speeds this pattern is reversed.  While the raw human data is hard to interpret, both data
and model show a shallow but consistent increase in skewness as speed increases.

Predictions for individual subjects reveals key variation in model behavior that is lost when we look at predictions for
an average subject.  Timing predictions for Subject 1S show a strong contraction in the timing bias at higher speeds,
quite similar to the experimental data from Subjects 1 and 3.  Predictions for Subject 2S shows no such contraction, and
appears qualitatively similar to Subject 2 from our experiment. The difference in subject timing behavior at high
speeds can be accounted for by differences in the shape of the prior distribution.  Stocker and Simoncelli \cite{alan2006} show
that under some conditions the speed bias is proportional to the slope of the prior distribution in the normalized
logarithmic space. The slope of the prior for Subject 1S flattens out at faster speeds while the Subject 2S prior
maintains a constant slope across speeds (Figure ~\ref{fig:datsub}).  We see that this difference in prior is reflected
by a contraction in timing bias for Subject 1S and no contraction in bias for Subject 2S. This suggests that the
qualitative differences in human timing response at high speeds can be partially explained by differences in the prior
shape.

\section{Discussion}
Experimentally, we show that the contrast speed visual illusions is reflected in timing actions.  To support our
conclusion we constructed an interception model grounded in a detailed Bayesian model for speed and showed that our
model produces similar biases in timing actions. We argued that differences in timing biases across subjects can be
partially explained by differences in the shape of the prior distribution, further strengthening the link between
Bayesian perception on the one hand and actions on the other.  The qualitative link between the model variance and
subject variance and the similar upward trend in skewness shows that most of the behavioral noise is explained by the
noise parameters in our model: the motor noise that grows proportionally with timing magnitude and the perceptual noise
due to contrast.  These results challenge the notion that visual illusions are not reflected in action
\cite{GooddaleREV} and demands a more nuanced view.  Our results suggest that visual illusions that can be
explained in terms of Bayesian inference will similarly affect action.

Our modeling framework for the interception task draws inspiration from recent Bayesian models of human intuitive
physics \cite{Battaglia11, Sanborn09}.  These models combine uncertainty over physical variables, based on perception
and prior knowledge, with a deterministic Newtonian physics model to infer the most likely behavior of the physical
world.  We expand on this work by investigating a simple speed timing task in detail and thus show that biases in one
component of a physics model can percolate through to affect inference about other physical variables, such as time.  In
future work we intend to teach subjects novel physical relationships and see if speed biases continue to influence
judgments about other physical variables.

Over all, we see this work as part of an ongoing effort to probe the degree to which Bayesian models generalize across different behavioral
tasks.  Our results indicate that the Bayesian model for speed perception generalizes to novel task domains that are different from the one
it was originally conceived and that the speed prior transfers across these domains.

\subsection*{Acknowledgments}
We thank all the subjects who participated in the experiments. This work was made possible by the Office of Naval Research (grant
N00014-11-1-0744).

\bibliographystyle{unsrt}
\bibliography{mybib}

\end{document}